\documentclass[12pt]{article}
\usepackage{amssymb, latexsym, epsfig}
\newcommand{\newsection}[1]{
\vspace{10mm}
\pagebreak[3]
\refstepcounter{section}
\setcounter{equation}{0}
\setcounter{subsection}{0}
\setcounter{footnote}{0}
\addcontentsline{toc}{section}{\protect\numberline{\arabic{section}}{{\rm #1}}}
\begin{center}
{\large \sc \thesection. #1}
\end{center}
\nopagebreak
\medskip
\nopagebreak}
\newcommand{\newsubsection}[1]{
\vspace{5mm}
\pagebreak[3]
\refstepcounter{subsection}
\addcontentsline{toc}{subsection}{\protect
\numberline{\arabic{section}.\arabic{subsection}}{#1}}
\noindent{ \sc \thesubsection. #1}                 
\nopagebreak
\vspace{2mm}
\nopagebreak}

\newcommand{\bye}{\end{document}}
\newcommand{\bq}{\begin{quote}}
\newcommand{\eq}{\end{quote}}

\renewcommand{\theequation}{\thesection.\arabic{equation}}

\newcommand{\ben}{\begin{enumerate}}
\newcommand{\een}{\end{enumerate}}
\newlength{\extraspace}
\newlength{\extraspaces}
\newcounter{dummy}
\newcommand{\bc}{\begin{center}}
\newcommand{\ec}{\end{center}}
\newcommand{\be}{\begin{equation}
\addtolength{\abovedisplayskip}{\extraspaces}
\addtolength{\belowdisplayskip}{\extraspaces}
\addtolength{\abovedisplayshortskip}{\extraspace}
\addtolength{\belowdisplayshortskip}{\extraspace}}
\newcommand{\ee}{\end{equation}}

\newcommand{\ba}{\begin{eqnarray}
\addtolength{\abovedisplayskip}{\extraspaces}
\addtolength{\belowdisplayskip}{\extraspaces}
\addtolength{\abovedisplayshortskip}{\extraspace}
\addtolength{\belowdisplayshortskip}{\extraspace}}
\newcommand{\ea}{\end{eqnarray}}
\newcommand{\nonu}{\nonumber \\[.5mm]}
\newcommand{\is}{& \!\! = \!\! &}
\newcommand{\ban}{\begin{eqnarray*}
\addtolength{\abovedisplayskip}{\extraspaces}
\addtolength{\belowdisplayskip}{\extraspaces}
\addtolength{\abovedisplayshortskip}{\extraspace}
\addtolength{\belowdisplayshortskip}{\extraspace}}
\newcommand{\ean}{\end{eqnarray*}}
\newcommand{\baa}{                         
\addtocounter{equation}{1}
\setcounter{dummy}{\value{equation}}
\setcounter{equation}{0}
\renewcommand{\theequation}{\thesection.\arabic{dummy}\alph{equation}}
\begin{eqnarray}
\addtolength{\abovedisplayskip}{\extraspaces}
\addtolength{\belowdisplayskip}{\extraspaces}
\addtolength{\abovedisplayshortskip}{\extraspace}
\addtolength{\belowdisplayshortskip}{\extraspace}}
\newcommand{\eaa}{                                       
\end{eqnarray}
\setcounter{equation}{\value{dummy}}
\renewcommand{\theequation}{\thesection.\arabic{equation}}}

\renewcommand{\d}{{{\partial}}}
\newcommand{\half}{{\textstyle{1\over 2}}}

\def\Bbb{\bf}

\newcommand{\Z}{{\Bbb Z}}
\newcommand{\R}{{\Bbb R}}

\newcommand{\C}{{\Bbb C}}

\newcommand{\inv}{^{-1}}

\newcommand{\delbar}{\overline{\partial}}
\newcommand{\dbar}{{\overline{\partial}}}

\newcommand{\Cbar}{{\overline{C}}}

\newcommand{\ibar}{{\overline{\imath}}}
\newcommand{\Ibar}{{\overline{I}}}
\newcommand{\jbar}{{\overline{\jmath}}}

\newcommand{\kbar}{{\overline{k}}}

\newcommand{\tbar}{{\overline{t}}}
\newcommand{\wbar}{{\overline{w}}}

\newcommand{\xbar}{{\overline{x}}}
                        
\newcommand{\zbar}{{\overline{z}}}

\newcommand{\Psibar}{{\overline{\Psi}}}

\newcommand{\taubar}{{\overline{\tau}}}

\newcommand{\cM}{{\cal M}}
\newcommand{\cF}{{\cal F}}
\newcommand{\cN}{{\cal N}}

\renewcommand{\Im}{{\rm Im\,}}

\newcommand{\del}{\partial}
\def\a{\alpha} 
\def\b{\beta} 
 
\def\G{\Gamma}

\def\w{\omega}
\def\W{\Omega}

\def\<{\langle}
\def\>{\rangle}

\newfont{\gothic}{eufm10 scaled\magstep1}

\newcommand{\hhbar}{{\overline{h}}}

\newcommand{\hM}{{\widehat{\cal M}}}

\newcommand{\omegabar}{{\overline{\omega}}}
\newcommand{\sk}{}
\newcommand{\ret}{\nonumber \\}
\renewcommand{\d}{\partial}

\newcommand{\cNbar}{\overline{\cN}}
\newcommand{\cFbar}{\overline{\cF}}
\newcommand{\bea}{\begin{eqnarray}}

\newcommand{\bZ}{\mathbb Z}
\newcommand{\eea}{\end{eqnarray}}
\newcommand{\ga}{\alpha}
\newcommand{\gb}{\beta}
\newcommand{\gd}{\delta}
\newcommand{\gD}{\Delta}
\newcommand{\gS}{\Sigma}

\newcommand{\gG}{\Gamma}
\newcommand{\gl}{\lambda}
\newcommand{\go}{\omega}
\newcommand{\gO}{\Omega}
\newcommand{\gT}{\Theta} 
\newcommand{\gt}{\theta} 
\newcommand{\gs}{\sigma} 
\newcommand{\gtbar}{\overline{\theta}} 
\newcommand{\Omegabar}{\overline{\Omega}}
\newcommand{\Psihat}{\hat{\Psi}}
\newcommand{\gatilde}{\tilde{\alpha}}
\newcommand{\gobar}{\overline{\omega}}
\newcommand{\gohat}{\hat{\omega}}
\newcommand{\zhat}{\hat{z}}

\begin{document}
\begin{titlepage}
\begin{center}

{\hbox to\hsize{\hfill PUPT-2030}}
{\hbox to\hsize{\hfill SPIN-2002/10}}
{\hbox to\hsize{\hfill ITFA-2002-9}}

\vspace{3\baselineskip}

{\Large \sc On The Partition Sum of The NS Five-Brane}

\vspace{1.5cm}

{ \sc Robbert Dijkgraaf${}^{1,2}$, Erik Verlinde${}^{3}$ and 
Marcel Vonk${}^{4,1}$} \\[1cm]
{${}^1$ \it Institute for Theoretical Physics, University of Amsterdam,\\ 
1018 XE Amsterdam, The Netherlands}\\[3mm]
{${}^2$ \it Korteweg-de Vries Institute for Mathematics, 
University of Amsterdam,\\ 
Plantage Muidergracht 24, 1018 TV Amsterdam, The Netherlands}\\[3mm]
{${}^3$ \it Physics Department, Princeton University, Princeton, NJ 
08544, USA}\\[3mm]
{${}^4$ \it Spinoza Institute, University of Utrecht,\\  Leuvenlaan 4, 
3584 CE Utrecht, The Netherlands}\\[.5cm]

\vspace*{1cm}
\large{

{\sc Abstract}}\\
\end{center}

\begin{quote}
We study the Type IIA NS five-brane wrapped on a Calabi-Yau manifold
$X$ in a double-scaled decoupling limit. We calculate the euclidean
partition function in the presence of a flat RR 3-form field. The
classical contribution is given by a sum over fluxes of the self-dual
tensor field which reduces to a theta-function. The quantum
contributions are computed using a T-dual IIB background where the
five-branes are replaced by an ALE singularity. Using the supergravity
effective action we find that the loop corrections to the free energy
are given by B-model topological string amplitudes. This seems to
provide a direct link between the double-scaled little strings on the
five-brane worldvolume and topological strings. Both the classical and
quantum contributions to the partition function satisfy (conjugate)
holomorphic anomaly equations, which explains an observation of Witten
relating topological string theory to the quantization of three-form
fields.

\end{quote}

\noindent

\end{titlepage}

\newpage
\newsection{Introduction and Summary} 
The type IIA Neveu-Schwarz five-brane remains one of the more
mysterious objects in string theory. Its world-volume theory is
described by six-dimensional little strings \cite{Dijkgraaf:1996bs,
Dijkgraaf:1996bq, Berkooz:1997md, Seiberg:1997nt, Losev:1997mm} that
in the low-energy limit reduce to a (2,0) superconformal field theory
which contains self-dual two-form fields. One of the problems of the
five-brane world-volume theory is that it is in general quite hard to
work with self-dual tensor fields since they do not allow a
straightforward covariant Lagrangian formulation. A second problem,
the fact that little string theory did not seem to have a natural weak
coupling limit, was illuminated by Giveon and Kutasov in
\cite{Giveon:1999ds1, Giveon:1999ds2}, where they found a double scaling
limit in which the world-volume theory decouples from the space-time
theory and moreover a weak coupling expansion exists. In this paper we
investigate the type IIA NS five-brane in a similar limit that
involves the background RR three-form field. The concrete aim of this
paper will be to calculate the quantum corrected euclidean partition
function of the five-brane, wrapped on a Calabi-Yau manifold $X$ in
the presence of a flat RR 3-form field $C$ on $X$.

\sk The motivation for this calculation is twofold. First of all a
computation of the partition function allows us to learn more about
the little string theory itself. Since a wrapped brane has a natural
twisted supersymmetry algebra the partition function can be viewed as a
supersymmetric index \cite{Bershadsky:1995qy} and as such it decouples
from certain moduli. Indeed we will see that it only depends on the
complex structure moduli (or equivalently, conformal structure) and RR
background of $X$. That is, the five-brane partition function only
depends on the IIA hypermultiplet moduli of the Calabi-Yau.

\sk This last observation is closely related to a second application of
our computation. In type IIA Calabi-Yau compactifications there are
possible five-brane instanton contributions to the hypermultiplet
metric and other four-dimensional quantities coming from euclidean NS
five-branes wrapping the CY \cite{Becker:1995kb}. In the low energy
decoupling limit these contributions are represented by the five-brane
partition function coupled to the RR background field. So our
computation could be helpful in determining these instanton
corrections.

\sk We will find that the free energy $F$ can be expressed as a
semiclassical expansion in the effective string coupling constant. The
classical contribution is given by the sum over fluxes of the
self-dual tensor field that lives on the five-brane. This result is
familiar \cite{Witten:1997hc} (see also e.g.\
\cite{Dolan:1998qk,Henningson:1999dm,Henningson:2001wh}) and 
can be expressed as a theta-function,
\be
 Z_X^{cl} = \overline{\Theta_{\a,\b}(x;z)} 
 \ee
which depends on the fluxes $x$ of the $C$-field, on the coordinates $z$ 
of the complex structure moduli space of $X$, and on a choice of a kind of 
``spin structure'' on $X$ given by $\ga, \gb$.

\sk We claim however that in this particular double-scaled decoupling
limit there are non-trivial quantum corrections, even in the case of a
single five-brane due to the presence of the background $C$ field.
The crucial ingredient in deriving the quantum part of the result will
be the fact that we can change to a T-dual point of view where we can
deduce the exact quantum contributions. It is known
\cite{Ooguri:1995bh} that a configuration of $k$ euclidean NS
five-branes in Type IIA string theory compactified on $X \times \R^4$
can be T-dualized to a Type IIB compactification without five-branes
but with a non-trivial metric of the form $X \times M^4$. Here $M^4$
is an ALE background of type $A_{k-1}$ also known as a gravitational
instanton. Since we go from the IIA to the IIB theory with a T-duality
transverse to the Calabi-Yau the compactification $X$ does not change
--- this in contrast with mirror symmetry.

\sk We will argue that in our limit, the relevant terms in the low-energy
supergravity description of this system will be the so-called F-terms
(BPS saturated terms) that can be computed from B-model topological
string amplitudes. These terms have the form
\be
 S =  \sum_{g>0}\int_{M^4} \cF_g(z,\zbar) T^{2g-2} R_- \wedge R_-,
\ee
where $T$ is the so-called graviphoton field and $R_-$ the
anti-self-dual part of the Riemann curvature. On an ALE singularity
the curvature term gives a localized $\delta$-function contribution
that corresponds to the decoupled five-brane quantum effective action
in the IIA picture.  So we are led to the conclusion that quantum
partition function is given by
\be
 Z_X^{qu} = \exp \sum_{g>0} \cF_g(z,\zbar) \gl^{2g-2},
\ee
where $\gl$ is the effective string coupling constant proportional to
the expectation value of $T$. The full answer for the partition
function,
\be
 Z_X = Z_X^{cl} \cdot Z_X^{qu},
\ee
is then obtained by combining the classical and quantum contributions.

\sk As we will explain in more detail later, in the simple case of $X=K3
\times T^2$, where by supersymmetry the higher loop terms $\cF_{g>1}$
vanish, this gives the familiar result
\be
 Z_{K3\times T^2} = {\theta_{4,20}(\tau,\bar\tau) \over \eta(\tau)^{24}},
\ee
where $\tau$ is the modulus of the torus. In this expression, the Narain 
theta-function comes from the classical sum over fluxes and the 
eta-function results from the $\cF_1$ one-loop term. So, in some sense our 
results suggest that this form of ``theta/eta'', so familiar from chiral 
conformal field theories in two dimensions, generalizes naturally to 
Calabi-Yau manifolds.

\sk Remarkably the classical and the quantum part of the partition
function both satisfy (actually conjugated) holomorphic anomaly
equations. This anomaly equation was first derived by by Bershadsky,
Cecotti, Ooguri and Vafa \cite{Bershadsky:1993ta, Bershadsky:1994cx}
in the case of the topological string amplitudes. Subsequently Witten
\cite{Witten:1993ed} observed that a similar relation can be obtained
by quantizing the space of flat three-forms $H^3(X;\R)$.  In the
latter case the corresponding wave function is naturally given by the
theta-function $\Theta$ which appears here as the classical
contribution of the five-brane partition function. The two
contributions can thus be regarded to take values in dual line bundles
over the moduli space of complex structures on $X$.

\sk
This paper is organized as follows. We begin with a description of our
system and the limit we take in section \ref{sec:system}. In section
\ref{sec:classpf} we evaluate the classical part of the
partition function for a single NS five-brane in type IIA string
theory. We pay particular attention to the fact that the two-form
field appearing in the action has a self-dual field strength, using
methods developed by Witten \cite{Witten:1997hc}. We also consider two
specific examples to clarify our results. The first is an NS
five-brane wrapped on a $K3 \times T^2$ manifold; the second example
considers a $\bZ_2$-orbifold of this \cite{Borcea:1997ms} which in
most respects --- though not in all, as we will discuss --- is a
Calabi-Yau manifold. After this, we describe the quantum contributions
to the partition function in section \ref{sec:quantumpf} --- now with
an arbitrary number of five-branes in a Coulomb phase. We cannot
really derive these contributions from our IIA point of view, so here
we change to the T-dual description. We first check that the classical
result in this picture agrees with the one we obtained on the IIA
side, and then derive the expression for the quantum
contributions. After this we come back to the two examples we studied
in section \ref{sec:classpf}. In section \ref{sec:remarks}, we make
some interesting remarks and state some open questions about our
results.  Section \ref{sec:conclusion} contains our conclusions, and
appendix \ref{app:bcov} contains the more detailed calculations that
we need in order to derive the holomorphic anomaly equation in section
\ref{sec:remarks}.

\newsection{Description of the system}
\label{sec:system}
The system we study in this paper has two T-dual descriptions. We will
begin by describing the system from a type IIA point of view, and then
explain what the same system looks like if we T-dualize it to a type IIB
system.

\newsubsection{The type IIA picture}

\label{sec:systemIIA} \noindent
We compactify euclidean type IIA string theory on a fixed Calabi-Yau
threefold $X$. In this theory, we put $k$ euclidean NS five-brane
instantons that are completely wrapped around the CY and hence are
point-like in the remaining four space-time directions. We denote the
generic distance between these five-branes in four-dimensional space
by $\gD x$. Following Giveon and Kutasov \cite{Giveon:1999ds1,
Giveon:1999ds2}, we define a mass scale
\be
 \label{eq:limit}
 m_W \equiv \frac{\gD x}{g_s^A l_s}.
\ee
Here $g_s^A, l_s$ denote the usual type IIA string coupling constant and
the string length. Originally, this quantity was defined for NS
five-branes in IIB string theory where it can be interpreted as the
mass (in units of the string mass) of a ``W-boson'' formed by stretching a
D-string between two of the five-branes. Of course in IIA theory there are
no D-strings, but we can for example interpret this quantity as the mass
density of stretching D2-branes in string units. We want to take a
limit where $\gD x \to 0$, but the five-branes can still be viewed as
independent objects in a Coulomb phase. This can be done, as was argued in 
\cite{Giveon:1999ds1, Giveon:1999ds2}, by sending $g_s^A \to 0$ as well,
in such a way that $m_W$ is large. In fact, as we shall explain below, we
want to take a limit where $m_W \to \infty$.

\sk Since $g_s^A$ goes to zero, the theory on the NS five-branes
decouples from the bulk string theory, and we obtain little string
theory. (Besides the references mentioned in the introduction, we
refer to \cite{Aharony:1999ls} for a review of little string theory.)
Moreover, Giveon and Kutasov argue that this theory becomes weakly
coupled if we take $m_W$ large. Of course, since the masses of the
objects that stretch between the five-branes become large, the
theories on the different five-branes will also decouple from each
other. This means we are indeed in a Coulomb phase, and we will use
this fact later on to argue that the partition function of the full
system can be written as the $k$-th power of the partition function of
a single five-brane.

\sk The low-energy theory on the five-brane contains a self-dual
two-form field, five scalar fields and the fermions needed for
supersymmetry. Of the bosonic fields, the self-dual two-form will be
the only one that we consider in this paper. Four of the scalars
correspond to the position of the five-brane, which in our setup is
fixed. We should really take the fifth scalar $Y$, which corresponds
to the position of the five-brane on the M-theory circle, into
account. However, note that we will generally assume that the
Calabi-Yau has strict $SU(3)$ holonomy and therefore does not have any
non-trivial 1-cycles. So the classical solutions of the field
equations for $Y$ cannot have any winding number, and only the
constant field configuration contributes\footnote{In our example of
$K3 \times T^2$, there are 1-cycles, and we will have to take the
winding numbers of $Y$ into account.}. 

%There can of course be quantum
%contributions coming from $Y$, but we will not be considering these in
%this paper. Finally, we also ignore the fermion fields in our
%discussion; their contributions can be included using supersymmetry
%arguments.

\sk
We will also turn on a background {\it flat} Ramond-Ramond three-form
field $C$ which along the directions of $X$ will have $dC=0$. This
field can have fluxes through three-cycles of the Calabi-Yau, and it
will couple to the field strength $H=dB$ of the self-dual two-form
field living on the five-branes. Technically it is classified by an
element of $H^3(X,U(1))$. (Extra subtleties in the quantization coming
from the K-theory interpretation of RR fields are absent for CY
three-folds, since there is no discrete torsion.)

\newsubsection{The type IIB picture}

\label{sec:systemIIB} \noindent
When we compactify the system described above on an extra circle, we can
do a T-duality in this direction. It is known \cite{Ooguri:1995bh} that
after such a T-duality, the five-branes disappear, but the geometry of the
resulting space becomes non-trivial and we obtain a Taub-NUT space. If we
let the compactification circle grow to infinite size on the IIB-side,
this space degenerates into an ALE-space of type $A_{k-1}$, where $k$ is 
the number of five-branes on the IIA-side. There is a problem with the 
case $k=1$ here, which we will discuss in section \ref{sec:missingfb}.

\sk
Now let us investigate what the limit described in the previous section 
corresponds to on the type IIB-side. After a T-duality, the string
coupling constant is
\be
 g_s^B = g_s^A \frac{l_s}{R_A},
\ee
where $R_A$ is the size of the compactifying circle. Now if we would have 
put our five-branes at fixed angles around the circle, $R_A$ would be 
proportional to $\gD x$, and we arrive at the result that the string 
coupling constant on the IIB-side is inversely proportional to $m_W$. In 
particular, it will go to zero in our limit. Note that this also implies 
that the four-dimensional Planck length
\be
 l_p = g_s^B l_s
\ee
is small compared to all other lengths in the IIB geometry, so that we
can trust the supergravity approximation to the IIB string theory.

\sk
Since the $C$-field in the IIA theory had all its legs along the
Calabi-Yau, after T-duality it will become a four-form $C_4$ which
from the four-dimensional point of view looks like a set of Abelian
one-form gauge fields --- where each three-cycle of the Calabi-Yau
gives rise to a single gauge field\footnote{More precisely, we should
take half of the cycles here, since a dual pair of cycles gives a
Hodge-dual pair of gauge fields related by an electromagnetic
duality.}.  The equations $dC=0$ of course have implications for these
gauge fields, which we will discuss in detail in section
\ref{sec:sugra}.

\sk
There is one special gauge field, the so-called graviphoton, which as
we shall see corresponds to the part of the four-form that multiplies
the holomorphic (3,0)-form of the Calabi Yau manifold. This field will
play an important role in our determination of the quantum
contributions to the five-brane partition function. On the ALE-space,
this field will have fluxes through the vanishing two-cycles, and an
overall flux $T_0$ through the two-sphere at infinity, which is the
boundary of the four-dimensional space without the T-duality
circle. We will be sending this overall flux (appropriately scaled by
powers of the string length) to infinity in such a way that the combination
\be
 \gl = g_s T_0
\ee
is held finite. We will see that $\gl$ is be the effective string coupling 
constant in our final result.

\newsection{The classical five-brane partition sum in type IIA}
\label{sec:classpf}
In this section, we want to calculate the classical contribution
(i.e.\ the contribution coming from solutions to the equations of
motion) to the five-brane partition function in the IIA setup we
described in the previous section. As we argued there, in the
particular limit we take we expect the partition function of $k$
five-branes to be simply given by the $k$-th power of the partition
function for one five-brane, so throughout this section we will
concentrate on a single five-brane. Note that from a four-dimensional
point of view, the sum over classical field configurations on the
five-brane is a sum over instanton quantum numbers, so we will use the
terminology ``partition sum'' and ``instanton sum'' interchangeably.

\sk
In section \ref{sec:classaction} we express the classical value of the 
action in terms of fluxes, while for the moment ignoring the self-duality
of the fields. In section \ref{sec:resum} we do a Poisson resummation
to obtain a factorized expression for the classical partition sum. The
calculation in these two sections is similar in spirit to the one in
\cite{Verlinde:1995ga}. In section \ref{sec:spinstr}, we follow Wittens
prescription \cite{Witten:1997hc} to incorporate the self-duality of the
tensor fields, and write down the final expression for the classical 
contribution to the partition sum. In section \ref{sec:introt}, we
rewrite this expression using an auxiliary variable $t$ that will be
useful to us later. Finally, in sections \ref{sec:examplek3t2} and
\ref{sec:examplebv} we use our results to study the examples of the
five-brane wrapped on $K3 \times T^2$ and $(K3 \times T^2)/\bZ_2$.

\newsubsection{The classical instanton action}

\label{sec:classaction}
\noindent
As described in section \ref{sec:systemIIA}, we are interested in the
action of a two-form field $B$ with field strength $H=dB$, coupled to a
background Ramond-Ramond three-form field $C$ with $dC=0$. In fact, using
the gauge invariance, this means that we can choose $C$ to be harmonic so
that $d^*C=0$ as well. The action is
\be
 \label{eq:action0}
 S = {1\over 4 \pi} \int_X \half (H-C)\wedge *(H-C) - i H \wedge C,
\ee
where $X$ denotes the Calabi-Yau manifold. According to the classical 
equation of motion $H$ satisfies $d^*(H-C)=0$. When we take our 
constraints on $C$ into account, this implies that $H$ is harmonic as 
well.

\sk
The numerical coefficients in the action (cf.\ \cite{Witten:1997hc,
Verlinde:1995ga}) are fixed by demanding that the fields can be made
self-dual and that the $C$-field only couples to the self-dual part of
$H$. In the presence of the background field, the condition of self-duality
reads
\be
 *(H-C)=-i(H-C),
\ee
where the factor $i$ comes from the fact that we work in euclidean
space. We will ignore this self-duality constraint for the moment, and
include it in our considerations in section \ref{sec:spinstr}.

\sk
The Calabi-Yau space has $b_3 = 2 h_{2,1}+2$ three-cycles through
which the field strength $H$ may have non-zero fluxes. We choose a
basis of canonical homology 3-cycles $A^I,B_I$ ($I=1,\ldots,g \equiv
h_{2,1}+1$) in $H_3(X,\Z)$, and denote the corresponding fluxes of $H$
by
\be
 \int_{A^I} H = 2\pi n^I,\qquad \int_{B_I}H=2\pi m_I,
\ee 
Similarly the periods of the background $C$-field will be denoted by
\be
 \int_{A^I} C = 2\pi x_1^I,\qquad \int_{B_I}C =2\pi x_{2,I}.
\ee
Note that since $H$ and $C$ are harmonic, every set of fluxes corresponds
to a unique solution of the equations of motion. Moreover, since $H$ is a
field strength there is the usual Dirac quantization condition saying that
$n^I, m_I$ have to be integers. The variables $x_1^I,x_{2,J}$ are angular
variables and take values in $\R/\Z$.

\sk
It will be useful to formulate the result using the notation of
special geometry which we will now briefly review. On the moduli space
$\cM_c$ of complex structures for the Calabi-Yau $X$, let us introduce the
homogeneous coordinates $z^I$ and the prepotential $\cF(z)$
(homogeneous of degree two) defined by the periods of the holomorphic
$(3,0)$ form $\Omega$:
\be
 \label{eq:prepot}
 \int_{A^I} \Omega = z^I, \qquad \int_{B_I} \Omega = \cF_I \equiv \del_I 
 \cF(z)
\ee
(with $\del_I \equiv \del/\del z_I$).
That the periods of $\gO$ over the $B$-cycles can indeed be written in 
this form is a standard result from special geometry. In terms of a dual 
canonical basis $\alpha_I,\beta^I$ of $H^3(X,\Z)$ this means we can write 
$\Omega = z^I\alpha_I + \del_I\cF(z) \beta^I$. The derivatives $\w_I = 
\del_I \W$ are clearly linearly independent, and --- since they correspond 
to first order perturbations of a (3,0)-form --- by Griffith's
transversality they span $H^{3,0}\oplus H^{2,1}$. Together, the forms
$\w_I$ and their complex conjugates $\bar{\w}_I$ form a basis of
$H^3(X,\C)$. The periods of the $\w_I$ are
\be
 \oint_{A^I} \omega_J = \delta^I{}_J, \qquad \int_{B_I} \omega_J = 
 \tau_{IJ}
\ee
where we introduced the period matrix\footnote{Note that --- as we will see
more clearly later --- the eigenvalues of the Hodge star have the
consequence that the imaginary part of $\tau_{IJ}$ is {\em not} positive
definite, but has signature $(h^{2,1},1)$.} 
\be
\tau_{IJ} \equiv \del_I 
\del_J\cF(z).
\ee

\sk
The field strength $H$ can now be written as 
\be
 H= 2\pi h^I \omega_I + 2\pi \hhbar^I \omegabar_I
\ee
with
\be
 \label{hmn}
 h^I = -{i\over 2}(m_J-\taubar_{JK}n^K) (\Im\!\tau\,\inv)^{IJ},
\ee
and $\bar{h}^I$ its complex conjugate. Similarly we can write
\be
 C = 2 \pi x^I\w_I + 2 \pi \xbar^I \omegabar_I
\ee
with
\be
 x^I= -{i\over 2} (x_{2,J}-\taubar_{JK}x_1^K) (\Im \tau \inv)^{IJ},
\ee
where again, following from the fact that the periods $x_1, x_2$ are real, 
$\bar{x}^I$ is the complex conjugate of $x^I$.

\sk
Some remarks about our conventions should be made here. In the following
all indices $I, J, K,\ldots$ are lowered with $\Im\tau_{IJ}$ and raised
with its inverse, which we have denoted above by $(\Im \tau \inv)^{IJ}$. 
Throughout this paper, we will interpret expressions of the form $\Im 
\tau^{-1}$ in this way, i.e.\ we first take the imaginary part and then 
the inverse.

\sk
To evaluate the classical action we make use of the fact that on a
Calabi-Yau manifold the action of the Hodge star on 3-forms is uniquely
determined by the complex structure, namely it acts as $(-1)^p i$ on
$(p,3-p)$-forms. Keeping this in mind, the evaluation of the action is now
straightforward. One finds 
\be
 \label{action}
 S_{m,n}(x; z) = 2 \pi \left\{ h^I\hhbar_I + x^I \xbar_I - 2 
 \xbar^I h_I -2{(h^I - x^I)\zbar_I (\hhbar^J - \xbar^J) z_J \over 
 |z|^2} \right\}
\ee
where $h^I$ depends on the integers $m_I$ and $n^I$ through its
definition (\ref{hmn}). Here and in the rest of the paper we use a
notation where $|z|^2 = z_I \zbar^I, z^2 = z_I z^I$ and $\zbar^2 =
\zbar_I \zbar^I$. The last term in $S_{m,n}$ arises because of the
fact that the Hodge star acts with a relative minus sign on the
$(3,0)$ part. Despite its appearance, this term is positive definite
since $|z|^2 < 0$. This point will be important when we introduce the
auxiliary variable $t$ in section \ref{sec:introt}. Note that to write
down the last term, we really need our manifold to be a strict
Calabi-Yau (i.e.\ it must have full $SU(3)$ holonomy) to make sure
that all three-forms are primitive so that there are no other
three-forms which have a relative minus sign under the Hodge star. In
our examples, we will see how to deal with slightly more general
cases.

\newsubsection{Poisson resummation of the instanton sum}

\label{sec:resum}
\noindent
The classical partition function for the full (i.e.\ not self-dual)
tensor field $H$ is given by a sum over all fluxes weighted by 
$e^{-S_{m,n}}$: 
\be
 \label{Z}
 Z(x;z)=\sum_{m,n}e^{-S_{m,n}(x;z)}
\ee
We are interested in evaluating the partition sum for the self-dual 
tensor field. In fact, the situation is similar to the situation for 
chiral bosons in two dimensions, see e.g.\ \cite{Witten:1997hc}.
There it was found that the partition sum for the unconstrained boson can 
be rewritten after a Poisson resummation as a sum of products of a chiral 
and an anti-chiral part. The partition function for the chiral boson is 
then the chiral factor of one of the terms, where the choice of a specific
term is equivalent to the choice of a spin structure on the 
two-dimensional manifold. We follow the same procedure here, and perform a 
Poisson resummation on the integers $m_I$:
\be
 \label{eq:zresummed}
 Z(x;z) = \sum_{p,n} e^{-\hat{S}_{p,n}(x;z)}
\ee
where $\exp(-\hat{S}_{p,n})$ is the Fourier transform of $\exp(-S_{m,n})$
with respect to the variable $m$.

\sk
To perform this Fourier transformation we need to invert the quadratic
form involving $m_I$ that appears in the action. We can read off this form
by putting $n^I=0$ and $x^I=0$ for a moment. This gives the quadratic term
\be
 {\pi \over 2} m_I \left\{ (\Im\tau\inv)^{IJ} - 2 {z^I\zbar^J\over 
 |z|^2} \right\} m_J.
\ee
To invert the matrix in brackets (or rather its symmetric part, which is 
what we need to do the gaussian integral), we use the identity
\be
 (\Im\tau\inv)^{IJ} - {z^I\zbar^J+\zbar^Iz^J\over |z|^2}= - (\Im {\cal 
 N}\inv)^{IJ}
\ee
with
\be
 \label{eq:defcaln}
 {\cal N}_{IJ}  =\taubar_{IJ} + 2i{z_I z_J\over z^2}
\ee
This identity is easily checked by multiplying both sides with $\Im
{\cal N}$. Note that in contrast with the holomorphic period matrix
$\tau_{IJ}$, the non-holomorphic matrix ${\cal N}_{IJ}$ does have a
positive definite imaginary part.

\sk
The Poisson resummation is now a tedious but straightforward calculation, 
and in the end we find
\be
 \label{eq:partfun}
 Z = (\det \Im \cN)^{1/2} \sum_{p_L, p_R} e^{-S_{p_L, p_R}},
\ee
where
\be
 S_{p_L, p_R} = - i \pi p_L^I \cNbar_{IJ} p_L^J - 2 \pi (2p_L - x - 
 \xbar)^I \xbar^+_I + i \pi p_R^I \cN_{IJ} p_R^J.
\ee
In this expression we used the notation
\be
 \xbar^+_I = (\Im \tau)_{IJ} \xbar^J - \frac{\zbar_I \zbar_J}{\zbar^2} (x +
 \xbar)^J,
\ee
and the sum is over the lattice
\bea
 p_L^I & = & \half n^I + p^I \ret
 p_R^I & = & \half n^I - p^I,
\eea
where $n^I$ and $p^I$ are the integers that appear in the sum 
(\ref{eq:zresummed}).

\newsubsection{The sum over spin structures}

\label{sec:spinstr}
\noindent
In analogy with the case of the chiral boson \cite{Witten:1997hc}, we 
would like to write this expression as a sum of products of a chiral and 
an anti-chiral term, where the sum is over spin structures on $X$. We 
can do this by observing that the $(p_L, p_R)$-lattice is a subset of the 
lattice of ``half-integers'' (i.e.\ integer multiples of $\half$)
satisfying the conditions
\bea
 p_L + p_R & = & n \ret
 p_L - p_R & = & 2p
\eea
for $p, n$ arbitrary integers\footnote{For notational convenience, we 
leave out the indices $I$ here.}. From the first equation we see that 
$p_L$ and $p_R$ are either both integer or both of the form ``integer + 
$\half$'', so we can replace the lattice sum by a sum over $k, l, \alpha$ 
where
\bea
 p_L & = & k + \ga \ret
 p_R & = & l + \ga
\eea
with $k,l$ integers, $k - l$ even and $\ga \in \{ 0, \half \}$. The
condition that $k - l$ is even can be satisfied by inserting a sum
over $\gb \in \{ 0, \half \}$ and a factor $ (-1)^{2 \gb (k - l)} $ in
the partition function. Now we have a sum over arbitrary integers
$k,m$, so we can factorize the partition function:
\be
 \label{eq:zchiralblocks}
 Z = (\det \Im \cN)^{1/2} \sum_{\ga^I, \gb_I \in \{ 0, \half \}} 
 \overline{\gT_{\ga, \gb}(x;z)} \gT_{\ga, \gb}(0;z)
\ee
with
\be
 \label{eq:classicalresult}
 \gT_{\ga, \gb}(x;z) = \sum_{(p - \ga)^I \in \bZ} \exp 2 \pi i \left\{ - \half 
 p^I \cN_{IJ} p^J + \gb_I p^I - i (2p - x - \xbar)^I x^+_I \right\}.
\ee
Since $\Im \cN$ is positive this theta sum converges. Note that it is
neither holomorphic in $x$ nor in $z$.
The sum over $\ga, \gb$ is the sum over spin structures we wanted to 
obtain, and we can now identify the classical partition sum of the 
self-dual tensor theory with one of the theta functions. That is, the
classical contribution to the five-brane partition function is
\be
Z_X^{cl} = \overline{\gT_{\ga, \gb}(x;z)}.
\ee
This is completely analogous to the way that chiral bosons in two
dimensions lead to partition functions that are equal to theta
functions depending on a choice of spin structure. The fact that we
have to choose such a spin structure can best be compared to a choice
of discrete theta-angles in the theory.

\sk
Note that from the way the background $C$ field (through its fluxes $x^I$) 
couples in (\ref{eq:zchiralblocks}) it follows that we have to take an 
{\em anti-holomorphic} theta function from this expansion. Ultimately, 
this is a consequence of the choice of the sign of the second term in
(\ref{eq:action0}). We made this choice in order to agree with the
existing literature, and in particular to have the matrix
$\cNbar_{IJ}$ appearing in our theta function. In section
\ref{sec:quantumpf}, we will see that if we choose the classical
result to be anti-holomorphic, the quantum corrections to the
partition function will have a {\em holomorphic} behaviour. The fact
that the two ingredients have conjugate behaviour is really a physical
result, but their individual behaviour is just a matter of convention.

\newsubsection{Introducing the auxiliary variable $t$}

\label{sec:introt}
\noindent
As will be explained in section \ref{sec:quantumpf}, the result for the
partition function can be understood from a T-dual perspective in terms of
the low energy $\cN=2$ supergravity theory in 4 dimensions coming from
the ten-dimensional type IIB theory. Indeed, the matrix ${\cal N}_{IJ}$ we
introduced in (\ref{eq:defcaln}) is known to represent the coupling of the
various $U(1)$ gauge fields in $\cN=2$ supergravity. In this context it is
also well-known that the coupling of the gauge fields can be expressed in
an $\cN=1$ superfield language in terms of the purely holomorphic
expression $\tau_{IJ}$ by introducing the Weyl multiplet that contains the
auxiliary graviphoton field $T$. Quite often it is useful to keep the
graviphoton as an independent field.

\sk
Similarly, in the present context it turns out to be useful to rewrite
the theta function (\ref{eq:classicalresult}) by introducing an
analogous auxiliary variable $t$.  One easily verifies that
\be
 \label{eq:deftheta}
 \Theta_{\a,\b}(x;z) = \sqrt{z^2} \sum_{(p-\ga)^I\in \Z} \int dt \,
  e^{-2\pi(x^I-2tz^I)\xbar_I} e^{2\pi i p^I\b_I} 
 \Psi_{p}(x;tz,\tbar \zbar)
\ee
where the ``conformal block'' $\Psi_p$ is defined through
\be
 \label{eq:confblock}
 \Psi_p(x;z,\zbar) = \exp 2\pi i\left\{-\half p^I \taubar_{IJ} p^J
 +2i p^I(z_I-x_I)+ i (z^I - x^I)(z_I - x_I)\right\}.
\ee
These conformal blocks are now {\it holomorphic} functions of $x$ but
have mixed $z$ and $\zbar$ dependence (the latter through the
anti-holomorphic period matrix $\taubar$).  Note that the variable $t$
only appears as a prefactor of $z^I$. Hence, it can effectively be
absorbed in a rescaling of the holomorphic $3$-form $\Omega$. The
function $\Psi_p$ is homogeneous of degree zero in the variable $\zbar$
and hence there is no $\tbar$-dependence in (\ref{eq:deftheta}).

\sk
It is instructive to write the anti-holomorphic conformal block, which as 
we saw is the one appearing in the partition function of the self-dual 
two-form, in the following way:
\be
 \overline{\Psi_p(0;z,\zbar)} = \exp 2 \pi i \left\{ \half (p^I + \zbar^I) 
 \tau_{IJ} (p^J + \zbar^J) - p^I \cFbar_I - \half \cFbar 
 \right\},
 \label{eq:genuszero}
\ee
where we set $x^I=0$, and $\cF$ is the prepotential defined in 
(\ref{eq:prepot}). The $z$-dependence is now all in the first term, 
and the last term can be viewed as the genus 0-term in a topological 
expansion. We will come back to this remark in section 
\ref{sec:anomalyeq}.

\sk
To further clarify the meaning of the variable $t$, let us briefly go 
back to the original expression (\ref{Z}) for $Z$ in terms of the 
action (\ref{action}). The partition sum $Z$ can almost be written as a 
gaussian integral of an auxiliary complex variable $t$, so that the last 
term in the action appears as the result of the gaussian integration. The 
``almost'' here refers to the fact that the quadratic term in the 
integrand actually has the wrong sign. Ignoring this for the moment and 
proceeding, we have
\be
 Z= |z|^2 \sum_{m,n} \int dt d\tbar \,  e^{-S_{m,n}(x; t,\tbar)},
\ee
where
\ba
 S_{m,n}(x; t,\tbar) \is 2\pi \left\{ h^I\hhbar_I + x^I \xbar_I - 2
 \xbar^I h_I \right.\nonu
 & & \ \ \ \left.+  2 t z^I (\hhbar_I - \xbar_I) + 2 \tbar \zbar^I
 (h_I - x_I)+  2 t\tbar z^I \zbar_I \right\}.
\ea 
Note that $|z|^2<0$, so the ``gaussian'' integral indeed has the wrong 
sign.

\sk
Using this action, the Poisson resummation is much easier to perform, and 
we find
\be
 Z=|z|^2 (\det \Im \tau) \sum_{(p_L,p_R)\in \Gamma} \int dt d\tbar \, 
 e^{-2\pi(\xbar^I -2 \tbar \zbar^I) x_I} \overline{\Psi_{p_L}(x;tz,\zbar)} 
 \Psi_{p_R}(0;tz,\zbar).
\ee
Introducing the sum over spin structures as before, we again find the 
result (\ref{eq:zchiralblocks}) with $\gT$ defined as in 
(\ref{eq:deftheta}). To see that the determinants match as well, note that
\be
 \det \Im \cN = \det \Im \tau \frac{|z|^2}{\sqrt{z^2 \zbar^2}}.
\ee
We have given an a posteriori justification for the introduction of the 
wrong-sign gaussian integral. This procedure can probably be made 
mathematically more rigorous by either introducing some cutoff at large 
$t$ or by viewing the gaussian integral as a formal expression meaning 
``evaluation of $t$ at the saddle points'', and showing that this commutes 
with the sum over $m,n$.

\newsubsection{Example: The classical partition sum on $K3 \times T^2$}

\label{sec:examplek3t2}
\noindent
Before writing down the full expression for the five-brane partition
function, we would like to stop here and consider two specific
examples. The first is the NS five-brane wrapped on a $K3 \times T^2$
manifold. Note that this is not a true Calabi-Yau manifold, since it has
$SU(2)$ holonomy group. Nevertheless, it will turn out that we can
describe it nicely using the techniques of this section. The nice thing
about this example is that the full answer --- including quantum
corrections --- is already known from two different points of view, so it
is a good way to test our results.

\sk
When we wrap a five-brane around $K3 \times T^2$, we can do a dimensional
reduction in two obvious ways: reduce on $K3$ or reduce on $T^2$. First of
all, when one reduces a single\footnote{In the case of $k$ five-branes the
partition function depends on the phase of the theory. In the ``Coulomb
phase'' the branes are separated in $\R^4$ and their contribution is
simply $Z_1^k$.  This corresponds to the dilute gas approximation. In the
``Higgs phase'' we expect a non-abelian $U(k)$ tensor theory. After the
reduction on $K3$  this phase should describe a bound state of $k$
heterotic strings, i.e. a ``long'' heterotic string that is wrapped $k$
times on $T^2$. By summing over all these wrappings (connected covers of
degree $k$) the partition function for $k$ branes is written as $Z_k = H_k
Z_1$ with $H_k$ the $k$-th Hecke operator.} five-brane on the $K3$
manifold, a two-dimensional CFT appears that coincides with the
world-sheet theory of the heterotic string compactified on $T^4$
\cite{Harvey:1995hs}. Since we use supersymmetric boundary conditions, the
resulting partition sum contains the contribution of only the left-moving 
oscillators. It is given by the well-known expression
\be
 \label{eq:pf204}
 Z_1 = {\theta_{4,20}(\tau,\bar\tau) \over \eta(\tau)^{24}},
\ee
where the denominator is the Dedekind $\eta$-function, and the numerator 
is the $\theta$-function on the momentum lattice $\Gamma^{4,20}$ for the 
heterotic string:
\be
 \label{eq:defthetafn}
 \theta_{4,20}(\tau,\bar\tau) = \sum_{(p_L,p_R)\in \G^{20,4}}
 q^{p_L^2/2} {\bar q}^{p_R^2/2},
\ee
with $\tau$ the complex modulus of the $T^2$, and $q=e^{2\pi i \tau}$.
Alternatively, when we reduce the theory on $T^2$, we obtain a $U(k)$ $\cN =
4$ super Yang-Mills theory. The partition function of the SYM theory on
$K3$ can be computed directly \cite{Vafa:1994tf,Minahan:1998vr} as an
exact sum of gauge instantons with the same result (\ref{eq:pf204}).

\sk
The result (\ref{eq:pf204}) consists of a classical result given by the
theta-function, and a quantum part given by the eta-function. One would
therefore expect to find the same classical contribution using the
techniques from this section. Let us now check that this is the case; we
will come back to the quantum contributions to this example in section
\ref{sec:examplesqc}. For the moment, we set the $C$-field to zero.

\sk
We start with some geometrical notions. For more on the geometry of
$K3$, the reader may for instance consult \cite{Aspinwall:1996k3}. Let us
denote the complex coordinate on $T^2$ by $w$, and its standard one-cycles
by $A, B$. The complex structure parameter of the torus is denoted by
$\tau$ as usual. As for $K3$, we have the standard integral basis of
two-cycles $\gS^I$ with an intersection matrix $M^{IJ}$ of signature
(3,19). The basis of two-forms that is dual to this is denoted by
$\eta_I$. Putting all of this together, we find a basis of three-forms on
$K3 \times T^2$ given by $\go_I = \eta_I \wedge dw, \gobar_I = \eta_I
\wedge d \wbar$, and a basis of three-cycles given by $A^I = \gS^I \times
A, B^I = \gS^I \times B$. If the holomorphic three-form on $K3$ is given
by $z^I \eta_I$, the holomorphic three-form on $K3 \times T^2$ is given by
$\gO = z^I \go_I$.

\sk
The main point in this calculation is to find the analogue for the $K3
\times T^2$ case of the matrix $\cN$ in (\ref{eq:defcaln}). The first term
is easily calculated, and is given by $\taubar_{IJ} = \taubar M_{IJ}$,
where $M_{IJ}$ is the inverse matrix of $M^{IJ}$. With the second term, we
have to be a bit careful. Note that the second term in (\ref{eq:defcaln})  
contains an operator that projects onto the $(3,0)$-direction. The origin
of this operator lies in the fact that on this direction, the Hodge star
acts with an opposite eigenvalue. However, on $K3$ there are three
directions in which the Hodge star has the ``wrong'' eigenvalue,
corresponding to the three negative eigenvalues of $M$. Therefore, the
result is that
\be
 \cN_{IJ} = \taubar M_{IJ} + 2 i (\Im \tau) M_{IK} {P^K}_J,
\ee
where $P$ now is a projection matrix on a three-dimensional subspace of
the momentum lattice $\Gamma_{3,19}$.

\sk
Finally, we must insert this expression for $\cN$ in our result 
(\ref{eq:zchiralblocks}), which without the $C$-field reads
\be
 Z = \sum_{\ga, \gb \in \{ 0, \half \}} \sum_{m,n \in \bZ - \ga} \exp 2
 \pi i \Big\{ \half m \cN m - \half n \cNbar n + \gb (m-n) \Big\}.
\ee
To see that this expression really can be expressed in terms of the 
$\theta$-function, choose a basis $p_I$ for $\Gamma_{3,19}$, so that 
$M_{IJ} = p_I \cdot p_J$. Now we call
\bea
 m^I p_I & = & \tilde{p}^L \ret
 n^I p_I & = & \tilde{p}^R,
\eea
and we decompose the $\tilde{p}^{L,R}$ into their space-like and time-like
parts with respect to this basis as
\be
 \tilde{p}^{L,R} = \tilde{p}^{L,R}_+ + \tilde{p}^{L,R}_-.
\ee
Using this notation, we see that
\bea
 m^I M_{IJ} m^J & = & \tilde{p}^L \cdot \tilde{p}^L \ret
 m^I M_{IJ} {P^J}_K m^K & = & \tilde{p}^L_+ \cdot
 \tilde{p}^L_+.
\eea
Collecting everything, we see that we can write the partition function as
\bea
 Z = \sum_{\ga, \gb \in \{ 0, \half \}}
 \sum_{\tilde{p}_L, \tilde{p}_R \in \Gamma_{3,19} + \ga}&& \!\!\!\!\!\!\!\!
\exp \pi i 
 \Big\{ \taubar (\tilde{p}^L_-)^2 + \tau (\tilde{p}^L_+)^2 - \tau
 (\tilde{p}^R_-)^2 - \taubar (\tilde{p}^R_+)^2 \ret
 && \qquad + (\Im \tau) 2 \gb (\tilde{p}^L - \tilde{p}^R) \Big\}
\eea
where $\Gamma_{3,19} + \ga$ is the momentum lattice translated by the
vector $\ga$. We see that the terms with $\tilde{p}^L$ and with $\tilde{p}^R$ nicely form two separate theta functions:
\be
 Z = \sum_{\ga, \gb \in \{ 0, \half \}} \gt \left[ \begin{array}{c} \ga
 \\ \gb \end{array} \right] (\tau) \, \gtbar \left[ \begin{array}{c} \ga \\ 
 \gb \end{array} \right] (\taubar),
\ee
where $\gt \left[ \begin{array}{c} \ga \\ \gb \end{array} \right]$ are the
theta functions on the lattice $\Gamma_{3,19}$ with characteristics $\ga,
\gb$. For $\ga = \gb = 0$, these theta functions reduce to (\ref{eq:defthetafn}). Finally, to get the partition function for the self-dual field, we have to take a single conformal block from this expression, and we obtain a single theta-function as our partition function.

\sk
Our result differs from (\ref{eq:pf204}) in two ways. First of all, we
find that we have to make a choice of spin structure and that the
theta-function we obtain can change according to this. This is precisely
what one would expect for a self-dual theory. Secondly, we obtain the
theta-function corresponding to a lattice of signature (3,19), whereas
there is a lattice of signature (4,20) in (\ref{eq:pf204}). This
difference is due to the fact that we only focussed on the holomorphic
two-form, but left the periodic scalar field on the NS five-brane out of
our discussion. As we noted in section \ref{sec:systemIIA}, this is
allowed for a Calabi-Yau manifold, but not on a manifold with nontrivial
one-cycles such as $K3 \times T^2$. Therefore, we should really add the
fluxes of the periodic scalar to our momentum lattice, in which case we  
would find a theta-function on the full lattice $\Gamma_{4,20}$.

\sk
Apart from these differences, which can be explained easily, we see that we
obtain exactly the correct classical partition function for this example,
so we made a successful consistency-check of our results so far. After we
discuss the quantum contributions for the general case in the next
section, we will see in section \ref{sec:examplesqc} that we can reproduce
the quantum factor in (\ref{eq:pf204}) as well. 

\sk
The inclusion of the $C$-field in our result is now straightforward, using
the full expression of formula (\ref{eq:zchiralblocks}). Since there is no
really elegant and meaningful expression for the the full result, we will
not write it down here.

\newsubsection{Example: The classical partition sum on $(K3\times
T^2)/\bZ_2$} 

\label{sec:examplebv}
\noindent
When the $K3$-manifold in the previous section is the double cover of a
so-called Enriques surface, we can divide out $K3 \times T^2$ by a certain
$\bZ_2$-action, which acts on $T^2$ as $w \to -w$, and has no fixed points
on $K3$. We will denote the resulting manifold by $\cM$. This type of
manifold was constructed by Borcea and Voisin \cite{Borcea:1997ms}, and
its application in string theory has been studied for instance in
\cite{Ferrara:1995sq, Aspinwall:1995dp, Harvey:1996tc}.

\sk
The three-cycles on $\cM$ can have two origins: there can be cycles that
are the projections of the invariant cycles on $K3 \times T^2$, and there
can be new ``torsion'' cycles $\gS$ for which $2 \gS = 0$. In
\cite{Aspinwall:1995dp}, it is shown that there are three of these torsion
cycles; they will not play a role in what follows, since they cannot
support fluxes of $H$ or $C$. The invariant three-cycles of $K3 \times
T^2$ can be constructed as follows. In \cite{Ferrara:1995sq}, the action
of the involution on the two-cycles of $K3$ is given: let $E_8$ be the
intersection matrix of the root lattice of the corresponding algebra, and
$\gs$ the intersection matrix of the hyperbolic lattice:
\be
 \gs = \left( \begin{array}{cc} 0 & 1 \\ 1 & 0 \end{array} \right).
\ee
We can choose a certain integral basis of $H_2(K3, \bZ)$ such that the 
intersection matrix is
\be
 E_8 \oplus \gs \oplus E_8 \oplus \gs \oplus \gs,
\ee
and $\bZ_2$ maps the first ten basis vectors on the next ten, and maps the
final two basis vectors to minus themselves. Therefore, we can construct
the invariant two-cycles as follows:
\be
 \begin{array}{lclclcl}
 A^i & = & A \times (\gS^i - \gS^{i+10}) & ~~~~~ &
 B_i & = & B \times M_{ij} (\gS^j - \gS^{j+10}) \\
 A^{11} & = & A \times \gS^{21} &&
 B_{11} & = & B \times \gS^{22} \\
 A^{12} & = & A \times \gS^{22} &&
 B_{12} & = & B \times \gS^{21},
 \end{array}
\ee
where $i=1, \ldots, 10$. The intersection numbers of these cycles are
\bea
 \label{eq:intersections}
 A^{i~} \cap B_{j~} & = & 2 \gd^i_j \ret
 A^{11} \cap B_{11} & = & 1 \ret
 A^{12} \cap B_{12} & = & 1,
\eea
and all other intersections vanish.

\sk
The story for the cohomology is similar: a basis of invariant three-forms
is
\bea
 \gohat_{i~} & = & \half(\go_i - \go_{i+10}) \ret
 \gohat_{11} & = & \go_{21} \ret
 \gohat_{12} & = & \go_{22}.
\eea
{F}rom \cite{Ferrara:1995sq}, we know that the holomorphic two-form of 
$K3$ obtains a minus sign under the $\bZ_2$, so the holomorphic three-form
$\gO_{\cM} = dz \wedge \gO_{K3}$ is invariant. In terms of the notation of
the previous section, this implies that $z^i = - z^{i+10}$, and $z^{11}$
and $z^{12}$ are arbitrary. The projected holomorphic three-form is
$\zhat^I \gohat_{I}$, where $I=1, \ldots, 12; \zhat^i = 2 z^i, \zhat^{11}
= z^{21}$ and $\zhat^{12} = z^{22}$.

\sk
The period matrix is now very easy to find; it is
\be
 \tau_{IJ} = \tau M_{IJ},
\ee
where $\tau$ is the modular parameter of our original torus and $M_{IJ} =
2 E_8 \oplus 2 \gs \oplus \gs$. The reader should note that in this
expression $2 E_8$ stands for the $E_8$ intersection form multiplied by
two, and not for $E_8 \oplus E_8$.

\sk
Note that the lattice $\Gamma_{2,10}$ corresponding to $M_{IJ}$ has two
timelike basis vectors, so in this respect we are not really dealing with
a Calabi-Yau yet. This is also reflected in the fact that the holonomy
group of $\cM$ is $SU(2) \times \bZ_2$ instead of the full $SU(3)$, and in
the fact that some of the intersection numbers in (\ref{eq:intersections})
are 2 instead of 1. On the other hand, $\cM$ is a real Calabi-Yau in the
sense that it does not have any free one-cycles and admits only one
covariantly constant spinor.

\sk
{F}rom this point on, our calculation is exactly the same as the one in 
the previous section, where we replace the lattice $\Gamma_{3,19}$ by
$\Gamma_{2,10}$ everywhere. The resulting classical partition function is
therefore the theta-function
\be
 Z_{\ga, \gb} = \gt \left[ \begin{array}{c} \ga \\ \gb \end{array} \right]
 (\tau),
\ee
on our lattice of signature (2,10) and for a specific choice $(\ga, \gb)$
of spin structure. Note that this is an automorphic function (i.e.\ it
has weight 0) on the lattice $\Gamma_{2,10}$, but since this lattice is
not self-dual, it is not a modular form under the full $SL(2, \bZ)$ acting
on $\tau$, but only under a subgroup of it.

\sk
As in our previous example, it is relatively straightforward to include
the $C$ field in our expressions, but since the resulting formula is not
very enlightening we will not present it here.

\newsection{IIB perspective and the quantum contributions}
\label{sec:quantumpf}
After having studied the classical contributions to the five-brane
partition sum, we would like to find an expression for the full partition
sum including quantum corrections. From the type IIA point of view, we
have not found a clear way to derive this expression. However, there is
some intuition, which we explain in section \ref{sec:quantumIIA}. Then
we turn to the T-dual description of our system (as described in section
\ref{sec:systemIIB}), where we can say much more. As we said before, we
find an $\cN = 2$ supergravity theory which we describe in section
\ref{sec:sugra}. Before finding the quantum result, in section
\ref{sec:classicalIIB} we explicitly check that this description gives us
the same classical partition sum as the IIA-description did. Then in
section \ref{sec:quantumIIB}, we find the quantum contributions and write
down the full partition function for our system. Section
\ref{sec:examplesqc} discusses the quantum aspects of our two examples.

\newsubsection{The quantum contributions --- type IIA point of view}

\noindent
\label{sec:quantumIIA}
As we mentioned before, our final claim in section \ref{sec:quantumIIB}
will be that the quantum corrections to the partition function are given
by the generating function of topological string amplitudes. From the IIA
point of view, we do not have a complete argument as to why only the
topological string amplitudes would appear in the partition function.
However, there might be some kind of ``twisting argument'' along the
following lines. Note that the geometry of the Calabi-Yau is such that
if we put type II string theory on it, the theory on the Calabi-Yau has
two remaining supersymmetries. Now when we introduce a five-brane, one of
these supersymmetries is broken, and we are left with $\cN = 1$. The
Lorentz group in six dimensions is $SO(6) \sim SU(4)$, whereas the
holonomy group of the Calabi-Yau is $SU(3)$. Therefore, there is a
remaining $U(1)$ inside the Lorentz group that we could twist with the
$U(1)$ in the supersymmetry algebra. By general arguments 
\cite{Bershadsky:1995qy} branes wrapped on non-trivial cycles have a twisted
supersymmetry algebra which causes them to be described by topological
field theories. For CY manifolds this twisting does not change the
physical properties, as e.g.\ is the case for the ${\cal N}=4$
super-Yang-Mills theory on a $K3$ manifold \cite{Vafa:1994tf}. This
might lead to a topological version of the little string theory that
is equivalent to the topological string.

\sk
Admittedly, this argument is very hand-waving and hard to make
precise. Therefore, we now turn to the T-dual description of our system,
where we have much better control over the quantum corrections to our
partition function, and we will really be able to see the appearance of
the generating function for topological string amplitudes.

\newsubsection{The field content of the supergravity theory} 

\label{sec:sugra}
\noindent
Now we turn to the T-dual description of our system, and we consider an
arbitrary number $k$ of five-branes. Compactifying ten-dimensional type II
supergravity on a CY manifold gives a four-dimensional $\cN = 2$
supergravity theory. Recall from section \ref{sec:systemIIB} that the
presence of the five-branes means we will be studying $\cN =2$
supergravity on an ALE-space of type $A_{k-1}$. If we start with IIA
supergravity, we obtain $h^{1,1}$ vector multiplets and $h^{2,1}$ 
hypermultiplets; if as in our case we start with IIB supergravity these
numbers are exchanged. Besides these multiplets, both theories contain the
universal hypermultiplet and the gravity (Weyl) multiplet.

\sk
On the IIA side, the complex moduli $z^I$ and the fluxes of the
$C$-field $x^I$ are the scalars in the $h^{2,1}+1$
hypermultiplets. Therefore, after T-duality, these fields (at least
the ones that are not in the universal hypermultiplet) must correspond
to fields in the $h^{2,1}$ vector multiplets on the IIB-side. Since
T-duality acts transversely to the CY, nothing changes to the complex
structure moduli $z^I$, so they are the scalars that appear in the
vector multiplets. (The overall length $|z|$ is the field that is in
the universal hypermultiplet.) On the other hand, the fields $x^I$
correspond to the three-form field $C$, which --- since it only has
legs along the CY --- corresponds to a four-form field $C^{(4)}$ on
the IIB side. After reduction on the CY this field turns into a set
of Abelian gauge fields with two-form field strengths $F^I$ --- one for
each three-cycle of the Calabi-Yau. The real components of $x^I$ are
most easily recovered when we realize that after reduction of the
gauge fields on the T-duality circle, they split up into a
three-dimensional real scalar and a three-dimensional gauge field
which can be dualized into a real scalar.

\sk
An important role in this section will be played by the auxiliary vector
field that is in the gravity multiplet, the graviphoton. As we will see,
its field equations will set it equal to the linear combination of $F^I$
that corresponds to the (3,0)-form of the Calabi-Yau.

\sk
Finally, we would like to see what the free parameters we summed over on
the IIA-side, i.e.\ the fluxes of the self-dual two-form, correspond to
on the IIB-side. The easiest way to see this is to look at the differences
of these fluxes on two different five-branes labeled, say, $a$ and
$b$. This difference is
\be
 p_a^I - p_b^I = \frac{1}{2 \pi} \left( \int_{A^I} H_a - \int_{A^I}
 H_b \right)
\ee
At first sight, this expression would seem to be zero, since we could draw
an interval $I$ between the two three-cycles on $a$ and $b$, and evaluate
this expression as the integral of $dH=0$ over $A^I \times I$. However,
note that the presence of the $C$-field changes this since a gauge
transformation $C \to C + dB$ is not really a symmetry in the presence of 
the five-branes, unless we compensate it by changing $H \to H +
dB$. Therefore, if $C_a$ and $C_b$ differ by such a gauge transformation,
so must $H_a$ and $H_b$, and the above expression becomes
\bea
 p_a^I - p_b^I & = & \frac{1}{2 \pi} \left( \int_{A^I} C_a - \int_{A^I}
 C_b \right) \ret
 & = & \frac{1}{2 \pi} \int_{A^I \times I} F^{(4)}
\eea
where $F^{(4)}$ is the field strength of $C$.

\sk
Now we can translate this expression to the IIB-picture. Note that if
we let the two five-branes $a$ and $b$ coincide on the IIA-side, this
corresponds to the vanishing of a specific two-cycle $S^2_{a,b}$ on the
IIB-side. Therefore, the interval $I$ is effectively replaced by this
two-cycle under T-duality. Of course, the four-form field strength also
becomes a five-form field strength, so our claim is that on the IIB-side
\bea
 p_a^I - p_b^I & = & \frac{1}{2 \pi} \int_{A^i \times S^2_{a,b}} F^{(5)}
 \ret
 & = & \frac{1}{2 \pi} \int_{S^2_{a,b}} F^I.
\eea
We therefore see that to relate our results of section \ref{sec:classpf}
to the IIB-picture, we will have to sum over the fluxes of the vector
fields through the vanishing cycles of the ALE-space. As we saw this only
gives us the differences of the fluxes --- we will come back to this point
later.

\newsubsection{The classical part of the partition function}

\noindent
\label{sec:classicalIIB}
As we argued in the previous section, we are interested in the terms
in the effective action that contain the gauge field strengths $F^I$
and the graviphoton field strength $T$. The terms in the supergravity
action containing these fields and the fields $z^I$ that are in the
same supermultiplets are \cite{deWit:1985ls}:
\be
 \label{eq:actiongrph}
 S = - \frac{1}{2 \pi}  \int \frac{i}{2} \tau_{IJ} F^I_- \wedge F^J_- - 2
 \zbar_I F^I_- \wedge T_- - \zbar^2 T_- \wedge T_-
\ee
The subscript on $F^I_-$ and $T_-$ refers to the fact that in our
setup these fields are anti-self-dual. In principle, the self-dual
complex conjugate terms are in the action as well, but we claim that
these fields do not contribute in our limit. This follows from the
fact that we send the distances $\gD x$ between the five-branes to
zero in the IIA-picture. In the IIB-picture, this means that the size
of the cycles supporting the self-dual fields goes to zero (in fact,
we know that when $\gD x = 0$, we are at the $A_{k-1}$-singularity,
where we only have the anti-self-dual cycles), and the fluxes of $F^I$
and $T$ only come from the anti-self-dual parts of these fields. We
are going to be a bit sloppy here, since we should really follow the
same procedure as in the preceding section --- first calculating
partition functions with arbitrary field strengths, and then taking
holomorphic roots, where we need to make a choice of spin
structure. We will not carry out this analysis again (the details are
left to the ambitious reader), but carry on in this more imprecise
way. We will see that for this reason, our results will lack some
phase factors, but for the rest they will be correct.

\sk
The field $T_-$ is an auxiliary field (the graviphoton field), so we may choose to integrate it out. This will turn out to correspond to integrating out the auxiliary $t$-parameter on the IIA-side. The equation of motion for $T_-$ is
\be
 T_- = - \frac{\zbar_I F^I_-}{\zbar^2},
\ee
and inserting this in the action we find
\bea
 S & = & - \frac{1}{2 \pi} \int \left( \frac{i}{2} \tau_{IJ} + \frac{\zbar_I
 \zbar_J}{\zbar^2} \right) F^I_- \wedge F^J_- \ret
 & = & - \frac{1}{2 \pi} \int \frac{i}{2} \cNbar_{IJ} F^I_- \wedge F^J_-,
\eea
where $\cN$ is the same matrix (\ref{eq:defcaln}) as before.

\sk
The solutions to the equation of motion for $F$ are harmonic two-forms,
and as we argued in the previous section, we will have to sum over the
fluxes of these harmonic forms. Therefore we expand $F_-^I$ in a basis
$\ga^a$ of harmonic forms: 
\be
 F^I_- = 2 \pi n^I_a \ga^a
\ee
The $2 \pi$ is inserted so that the flux quantization condition forces the
$n^I_a$ to be integers or half-integers, depending on a choice of spin
structure on our manifold which as we said we will leave out of our 
considerations. Since $F^I_-$ is anti-self-dual, we can take a basis where
the $\ga^a$ that appear in the above equation have $* \ga^a = - \ga^a$,
and hence the classical value of the action is
\bea
 S & = & - i \pi \cNbar_{IJ} n^I_a n^J_b \int \ga^a \wedge \ga^b \ret
 & = & - i \pi \cNbar_{IJ} n^I_a n^J_b Q^{ab}.
 \label{eq:actionnot}
\eea
We now turn to the question what the matrix $Q^{ab}$ is, or equivalently,
what the basis of harmonic forms $\ga^a$ is. If we would be on a compact
four-dimensional manifold with an $A_{k-1}$-singularity, the matrix $Q$
would correspond to the intersection matrix of the vanishing 2-cycles at
the singularity, which is the $A_{k-1}$ Cartan matrix corresponding to
the $SU(k)$-lattice:
\be
 Q^{ab} = \left( 
          \begin{array}{rrrrr}
            2 & -1 &  0 &  0 & \cdots \\
           -1 &  2 & -1 &  0 & \cdots \\
            0 & -1 &  2 & -1 & \cdots \\
            0 &  0 & -1 &  2 & \cdots \\
	            \vdots & \vdots & \vdots & \vdots & \ddots
          \end{array}
          \right)
\ee
The reason that we find this intersection matrix is that the overall
charge of the five-branes must be zero, since no flux can escape to
infinity. In that case, the basis vectors $\ga^a$ can be chosen to be of
the form $e^{a+1} - e^{a}$ (for $a = 1, \ldots, k-1$), where each of the
$e^i$ corresponds to a single unit of flux through the $i$-th
vanishing cycle. Note that since $e^i$ and $e^j$ have intersection
$\gd^{ij}$, these basis vectors indeed have the intersection matrix given
above. However, since we are on a noncompact manifold we can have a flux
through the sphere at infinity and hence a nonzero overall charge, so we
have to add a harmonic form $\ga^0$ corresponding to an overall unit of
flux through the sphere at infinity. Note that this two-form will not be
$L^2$-normalizable, whereas all the other basis vectors are.

\sk
Of course, we can also take a basis $\gatilde^a = e^a$ for $a=1, \ldots,
k$, in which case the matrix $Q^{ab}$ simply becomes the identity
matrix: $Q^{ab} = \gd^{ab}$. This is the Cartan matrix for $U(k)$ (or
$U(1)^k$), which is the correct group since we added a $U(1)$ degree of
freedom at infinity. The difference between $U(1)^k$ and $U(k)$ (or any
groups in between) cannot be seen in this analysis; the $U(1)^k$ case
corresponds to the Coulomb branch, where the five-branes are separated,
and the $U(k)$ case corresponds to the Higgs branch where the five-branes
are on top of each other. We will see that our analysis only covers the
Coulomb branch.

\sk
Inserting the matrix $Q^{ab} = \gd^{ab} $ in the action gives
\be
 S = - i \pi \sum_{a=1}^k n_a^I \cNbar_{IJ} n_a^J.
\ee
Now we have to perform the sum over the fluxes as we did on the IIA-side,
and we find
\bea
 Z^{cl}_{X, k} & = &  \sum_{ \{ n_a^I \} } \sum_{s.s.} (-1)^{s.s.} e^{i 
 \pi \sum_{a=1}^k n_a^I \cNbar_{IJ} n_a^J} \ret
 & = & \left( \sum_{ \{ n^I \} } \sum_{s.s.} (-1)^{s.s.} e^{i \pi n^I
 \cNbar_{IJ} n^J} \right)^k \ret
 & = & (Z_X^{cl})^k,
\eea
where the second sum stands for the sum over spin structures, and we
included a sign that depends on this spin structure and that should follow
from a careful analysis of the self-duality of the fields. $Z_1^{cl}$ is
indeed (modulo phase factors) the classical partition function of a single
five-brane we found in section \ref{sec:resum}. The final expression shows
that we are on the Coulomb branch, since the different five-branes do not
interact and all give the same contribution to the partition function. 

\sk
Similarly, we can do the calculation starting from (\ref{eq:actiongrph}),
and expanding the classical value of $T_-$:
\be
 T_- = 2 \pi \tbar_a \ga^a,
\ee
where we called the new variable $\tbar$ to agree with our previous result. The classical part of the partition function now contains a sum over the fluxes $n_a^I$  and an integral over the variables $\tbar_a$:
\bea
 Z^{cl}_{X,k} & = & \sum_{ \{ n_a^I \} } \int d\tbar_1 \cdots d\tbar_k 
 \exp 2 \pi \sum_{a=1}^k \left( \frac{i}{2} \tau_{IJ} n^I_a n^J_a - 2 
 \zbar_I n^I_a \tbar_a - \zbar^I \zbar_I \tbar_a^2 \right) \ret
 & = & \left( \sum_{ \{ n^I \} } \int d\tbar \exp 2 \pi \left[ \frac{i}{2}
 \tau_{IJ} n^I n^J - 2 \zbar_I n^I \tbar - \zbar^I \zbar_I \tbar^2 \right]
 \right)^k \ret
 & = & (Z_X^{cl})^k.
 \label{eq:actionwitht}
\eea
which is the result we found in section \ref{sec:introt} after we
introduced the variable $t$. Once again, we left out the sum over spin
structures and the corresponding signs, but of course they should be
included as well.

\newsubsection{The quantum contributions --- type IIB point of view}

\noindent
\label{sec:quantumIIB}
Now we want to find the quantum contributions to the five-brane partition
function. Therefore, we have to identify the terms in the 4-dimensional
supergravity action that come from these contributions. Recall that the
four-dimensional space is an ALE space. Such a space has a holonomy group
$SU(2) \subset SO(4) = SU(2) \times SU(2)$. In terms of the curvature,
this means that the self-dual part of the curvature $R = R_+ + R_-$
vanishes, so the curvature is purely anti-self-dual. The corrections to the
supergravity action that are non-zero when the self-dual part of the
curvature vanishes are explicitly known \cite{Antoniadis:1993ta}, and they
are of the form
\be
 S^{qu} = \int \sum_{g=1}^\infty R_- \wedge R_- (g_s T_-)^{2g-2}
 \cF_g(z,\zbar).  \label{eq:actionq}
\ee
Moreover, in \cite{Antoniadis:1993ta} it is explicitly shown that the
$\cF_g(z,\zbar)$ coincide with the topological string amplitudes at
genus $g$. We have indicated that in general through the mechanism of
the holomorphic anomaly\cite {Bershadsky:1994cx} these amplitudes also
have an anti-holomorphic dependence.
Note that when we take the limit described in section
\ref{sec:system}, the only component of $T$ that remains in this
expression is the component $T_0$ corresponding to the flux through
the sphere at infinity.

\sk
In our setup, the terms given above can easily be integrated. Note that
$(g_s T)^2$ is a harmonic four-form, so we can take it out of the integral
and replace it by the constant
\be
 \gl^2 = (g_s T_0)^2,
\ee
where in our limit $\gl$ takes over the role of the string coupling
constant. Moreover, since $R$ is anti-self-dual, $\int R_-^2 =
\int R^2 = \chi$, the Euler number of the ALE space. For an $A_{k-1}$
space this Euler number is $k-1$, and hence the contribution of these
terms to the partition function is
\bea
 e^{-S^{qu}} & = & e^{(k-1) \sum_{g \geq 1} \gl^{2g-2} 
 \cF_g(z,\zbar)} \ret & = & (Z^{qu}_X)^{k-1},
\eea
i.e.\ the $(k-1)$-th power of the quantum contribution that should 
correspond to a single five-brane on the IIA-side. Note that again, there
seems to be a ``missing five-brane'' here. We will say more about this in
section \ref{sec:missingfb}. Anyway this result leads to our main conclusion:
the quantum corrections in the particular decoupling limit of the five-brane
are given by the B-model topological string amplitudes
\be
Z^{qu}_X = \exp \left(\sum_{g\geq 1}\gl^{2g-2} \cF_g(z,\zbar)\right)
\label{quantum}.
\ee

\sk
So now, all the hard work is done, and we are ready to give our full
result. Combining the classical and the quantum results we found, the
formula for the single five-brane partition function in the limit we
considered is
\be
 \label{eq:finalresult}
 Z_X = \overline{\Theta_{\ga, \gb} (x;z, \zbar)} \exp \left( \sum_{g \geq 
 1} \cF_g (z, \zbar) \gl^{2g-2} \right),
\ee
where the first factor is the classical result (\ref{eq:classicalresult}),
and the second factor represents the quantum contributions.

\newsubsection{Examples: $K3 \times T^2$ and $(K3 \times T^2)/\bZ_2$}

\label{sec:examplesqc}
\noindent
We now return to the examples we studied in sections \ref{sec:examplek3t2}
and \ref{sec:examplebv}. The first of these was the example of the
five-brane wrapped around $K3 \times T^2$. We recall that the full
partition function in this case is known from two different calculations,
and reads
\be
 \label{eq:pf204b}
 Z_1 = \frac{\theta_{4,20}(\tau, \taubar)}{\eta(\tau)^{24}}.
\ee
In section \ref{sec:examplek3t2} we checked that our classical
contribution also gives a theta-function similar to the one above. We
would now like to study the quantum contributions to this partition
function.

\sk
In fact, because of the $\cN = 4$ supersymmetry there are only one-loop
contributions in this specific case. Therefore, to find the full quantum
corrected partition function, we need to know the one-loop amplitude for
topological strings on $K3 \times T^2$, which was calculated in
\cite{Harvey:1996fi}. The result is that
\be
 \cF_1^c =  -24 \log |\eta(\tau)|,
\ee
where the superscript $c$ refers to the fact that we take the contribution
to $\cF_1$ that depends on the complex structure moduli. Exponentialing
this, we find exactly the quantum contribution to (\ref{eq:pf204b}).

\sk
Next, we turn to our second example: $\cM = (K3 \times T^2)/\bZ_2$. Here,
as far as we know, we cannot compare our results to results in the
literature as we did in the $K3 \times T^2$ case. However, the topological
one-loop amplitude for this example is known, and it is calculated in 
\cite{Harvey:1996tc}. The result is that
\be
 \cF_1 = -\log |\eta^{24}(2 \tau) \Phi_{BE}(y)|,
\ee
where $y$ represents the holomorphic two-form (before dividing out the
$\bZ_2$) of $K3 \times T^2$, and $\Phi_{BE}$ is a certain modular form of
weight 4 on $\Gamma_{2,10}$ which was called the Borcherds-Enriques form in
\cite{Harvey:1996tc}. Putting this result together with the classical
result we found in section \ref{sec:examplebv}, we find that up to
one-loop corrections, the partition function for the five-brane on $\cM$
is
\be
 Z_X = \frac{\theta_{2,10}(\tau, \taubar)}{\eta^{24}(2 \tau) 
 \Phi_{BE}(y)}.
\ee
This result is still an automorphic form of weight 4 on $\Gamma_{2,10}$,
and transforms as a modular form under a subgroup of $SL(2, \bZ)$ acting
on $\tau$. Since $\cM$ only preserves half of the supersymmetry that $K3
\times T^2$ preserves, one expects that there are higher-loop
$\cF_g$-corrections to this expression as well.

\newsection{Remarks and open questions}
\label{sec:remarks}
The result we found in the previous section leads to many interesting
remarks and open questions. In this section, we mention three of these. We
begin by observing in section \ref{sec:anomalyeq} that both the classical
and the quantum part of the partition function satisfy a holomorphic
anomaly equation. Though we have some arguments as to why this is the
case, we do not yet understand the reasons for this completely. In section
\ref{sec:oneloop}, we ask some interesting questions about the one-loop
contribution to our partition sum. Finally, in section \ref{sec:missingfb}
we discuss the ``missing five-brane'', another point that is still not
completely clear to us.

\newsubsection{The holomorphic anomaly equation}

\label{sec:anomalyeq}
\noindent
We will now show that the holomorphic blocks $\Psi_p(x;z,\zbar)$
obtained in (\ref{eq:confblock}) satisfy exactly the same holomorphic
anomaly equations that the generating function for topological string
amplitudes satisfies, and that were found by Bershadsky et. al.
\cite{Bershadsky:1994cx} Indeed, it was noted by Witten
\cite{Witten:1993ed} that these equations can be interpreted as a
Knizhnik-Zamolodchikov type equation for the holomorphic quantization
of the space $H^3(X,\R)$ of three forms on the Calabi-Yau manifold
$X$. This space carries a natural symplectic structure $\int_X
\delta_1 \Omega\wedge \delta_2\Omega$, with $\delta_{1,2}\Omega\in
H^3(X,\R)$, which can be used to quantize it.

\sk
More precisely, a straightforward calculation shows that the functions
$\Psi_p(x;z,\zbar)$ obey the following equations:
\ba
 {\d\over \d z^I}\Psi_p(x;z,\zbar) \is \left[{\d\over\d x^I}-{i\over 2}
 C_{IJ}{}^K x^J {\d\over\d x^K} - {i\pi} C_{IJK}x^J
 x^K\right]\Psi_p(x;z,\zbar)\ret
 {\d \over \d\zbar^I}\Psi_p(x;z,\zbar)\is \left[{1 \over 16\pi i}
 \Cbar_I{}^{JK}{\d^2\over \d x^J\d x^K} + {1\over
 4i}\Cbar_{IJ}{}^J\right]\Psi_p(x;z,\zbar)
 \label{eq:largecfeq}
\ea
with 
\be
C_{IJK}=\d_K\tau_{IJ}=\d_I\d_J\d_K\cF
\ee
Note that if we put $x=0$ we find the equation
\be
 \label{eq:largepfeq}
 \left[{\d\over\d \zbar^I} - \dbar_I f_1\right] \Psi_p(0;z,\zbar)=
 \left[{1\over 16\pi i} \Cbar_I{}^{JK}{D^2\over D z^J
 D z^K}\right]\Psi_p(0;z,\zbar)
\ee
where the covariant derivatives are with respect to the metric
$\Im\tau_{IJ}$, i.e.\ they contain the Christoffel symbols
$\Gamma_{IJ}{}^K={i\over 2}C_{IJ}{}^K $.
The function $f_1(z)$ is defined by 
\be
 f_1 = -\half \log \det \Im \tau, 
\ee
so that $\delbar_I f_1 = \frac{1}{4i} {\bar{C}_{IJ}}^J = - \half 
{\bar{\Gamma}_{IJ}}^J$.

\sk
The reader who is familiar with the work \cite{Bershadsky:1994cx} will
note that (\ref{eq:largepfeq}) is very similar to the holomorphic
anomaly equation satisfied by the generating function for the
topological string amplitudes, and that (\ref{eq:largecfeq}) look like
the equations for the generating functions of correlation
functions. However, there are some differences between the equations
above and the equations found by Beshadsky et al. The reason for this
is that strictly speaking the variables $z^I$ are coordinates on the
``large phase space'' $\hM_c$ which parametrizes a complex structure
together with a choice of holomorphic $(3,0)$-form $\Omega$. In other
words they represent projective coordinates on the ``small phase
space'' $\cM_c$, the moduli space of complex structures.  In order to
relate our equations to the ones obtained by Bershadsky et al. and
Witten we therefore have to use the inhomogeneous coordinates on
$\cM_c$, that describe only the $(2,1)$-part variation of the
holomorphic 3-form.

\sk
We do this as follows. For given values of the moduli we can write for
any 3-form $x=x^I\w_I$
\be
 \label{eq:relbegin}
 x=x_\perp+ x_0,\qquad x_\perp\in H^{2,1},\ x_0\in H^{3,0}
\ee
with $(x_\perp,\zbar)=0$ and $x_0= z(x,\zbar)/(z,\zbar)$. Using this
split we can define a derivative $\d_I^\perp$ that describes the
variations in the $(2,1)$ direction satisfying $z^I\d_I^\perp = 0$. In
terms of the K\"ahler form
\be
 K=-\log(z,\zbar)=-\Im(\zbar^I\cF_I)
\ee
this derivative is given by
\be
 \d_I^\perp= \d_I+\d_IK z^J\d_J.
\ee
Moreover, in the topological string theory the amplitudes are defined on
the small phase space $\cM_c$, but they are in addition functions of the
string coupling constant $\lambda$, with the genus $g$ contribution
weighted by $\lambda^{2g-2}$. The function $\Psi_{p}(x;z,\zbar)$ can be
expressed in terms of these variables (as will be explained in the
appendix) by imposing the relations
\be
 \left(\lambda{\d\over\d\lambda}+x^I{\d\over\d x^I}+z^I{\d\over\d z^I}
 \right)\Psi=0,
\ee
and
\be
 \label{eq:relend}
 z^I{\d\over\d z^I}\Psi =z^I{\d\over\d x^I}\Psi,\qquad
 \zbar^I{\d\over\d \zbar^I}\Psi=0.
\ee
Using the relations (\ref{eq:relbegin}-\ref{eq:relend}) the BCOV equations
which are formulated on the small phase space are seen to coincide with
our equations as derived on the large phase space. Details of this
calculation are given in appendix \ref{app:bcov}.

\sk
Note that in our final answer (\ref{eq:finalresult}), the complex 
conjugate of $\Psi_p$ appears. This means that this factor in our answer 
satisfies an anti-holomorphic anomaly equation, whereas the quantum factor 
satisfies the (complex conjugate) holomorphic anomaly equation. Similarly, 
in (\ref{eq:genuszero}), $\cFbar$ appears, whereas in the quantum terms 
the holomorphic $\cF_g$'s appear\footnote{It is tempting to identify 
$\tbar$ with $\gl^{-1}$ to make this correspondence even more precise, but 
we do not have any clear physical arguments to support this idea.}. 
Following Wittens arguments in \cite{Witten:1993ed}, the anomaly equations 
state that the corresponding quantities live in a line bundle over the 
moduli space of complex structures, and are in some sense independent of 
the chosen background complex structure. Apparently both the classical 
result and the quantum contributions have this property separately. 
Moreover, if $\Psi_1$ and $\Psibar_2$ are two objects satisfying 
holomorphic and anti-holomorphic anomaly equations respectively, there is a 
natural inner product of the form
\be
 \int dx \, e^{-x \xbar} \Psi_1 \Psibar_2
\ee
which gives an anomaly-free result. In fact, this inner product is
background independent \cite{Witten:1993ed}. It has no $z$ and $\zbar$
dependence and is therefore a topological invariant associated to the
CY manifold. 

Why would one integrate over the $x$ variables? They are expectation
values of the supergravity scalar fields at infinity.  If we would
compactify the four-dimensional (euclidean) space-time $M^4$ one would
be forced to integrate over these vev's. The result then represents
the partition function of the full ten-dimensional Type II string on
the compact ten-manifold $X \times M$. Because of supersymmetry this
partition function corresponds to a topological index and is therefore
background independent.

\newsubsection{The one loop contribution}

\label{sec:oneloop}
\noindent
In section \ref{sec:examplesqc} we explicitly calculated the one-loop
corrections to the partition function in two examples --- where in one
case the one-loop correction actually was the only correction. Since
after the classical contribution the one-loop corrections are the
easiest ones to get a grip on, one might wonder if we could give some
more proof for our formulas by doing an explicit calculation. In fact,
a general expression for $\cF_1$ is known \cite{Bershadsky:1994cx,
Harvey:1996tc}, and it is given by
\bea
 \label{eq:toponeloop}
 \cF_1 & = & \sum_{0 \leq p,q \leq 3} pq (-1)^{p+q} \log {\det}'
 \gD^{p,q} \ret
 & = & 9 \log {\det}' \gD^{0,0} - 6 \log {\det}' \gD^{1,0} + \log {\det}
 \gD^{1,1},
\eea
where ${\det}' \gD^{p,q}$ is the determinant of the $\delbar$-operator
acting on $(p,q)$-forms with the zero-modes left out, and in the second line
we used the special properties of the Calabi-Yau. One would like to
reproduce this result directly from a one-loop calculation in the field
theory on the CY that we obtain in our limit. A naive guess would be that
the one-loop result would simply come from the determinants of the
Laplacians and Dirac operators for all the fields in our theory. However
(although again one has to be careful with the self-duality of the tensor
field), this does not seem to give the right numerical prefactors in
(\ref{eq:toponeloop}). The reason is probably related to the reason why
these exponents appear in (\ref{eq:toponeloop}): if we would calculate the
exact partition function for the topological string on a CY, the result
would be zero since there are fermion zero modes we integrate over. Here
the zero-modes have to be absorbed by introducing fermionic operators in
the path integral leading to the partition function. We suppose that a
similar calculation has to be done in the five-brane field theory, and one
would obtain the correct one-loop contribution (\ref{eq:toponeloop}). It
would be interesting to work this out in detail.

\newsubsection{The missing five-brane}

\label{sec:missingfb}
\noindent
At several points in this paper, we encountered a ``missing
five-brane'': we inserted $k$ five-branes in the IIA-theory, but the
results on the IIB-side seem to apply to $k-1$ five-branes. In section
\ref{sec:sugra}, this problem showed up in the form of a sum over
differences of fluxes instead of over the fluxes themselves. In section
\ref{sec:classicalIIB}, we could cure this problem, though in an
admittedly heuristic way, by introducing a $U(1)$ degree of freedom
corresponding to the overall flux at infinity. However, our quantum result
in section \ref{sec:quantumIIB} again showed only $k-1$ five-branes, and
we were not really able to cure this problem here.

\sk
There are two ways out of this problem. The nicest way would be to
completely understand the degrees of freedom at infinity, and include them
in our quantum calculation as well. However, one could also take the limit
of large $k$ and look at an ``ideal dilute gas of five-branes'', and only
look at the leading $k$-behaviour. This way of looking at things, though
of course much less satisfying theoretically, would clearly reproduce the
right results.

\sk
We feel that to completely solve this problem, a better analysis of the
T-duality between the set of five-branes in type IIA and the Taub-NUT
space, and especially of the decompactification limit, is needed. It would
be very interesting to carry out such an analysis.

\newsection{Conclusion}
\label{sec:conclusion}
In this paper, we computed the partition function for the NS
five-brane of type IIA string theory, including its quantum
corrections, in a double scaling limit very similar to
\cite{Giveon:1999ds1} in the presence of a flat RR three-form
field. The result is an expansion in the string coupling constant,
where the leading term is the sum over classical fluxes of the
self-dual two-form field that lives on the five-brane, and the
corrections are given by the higher-loop amplitudes of B-model
topological strings.

\sk
There are many interesting open questions, three of which we described in
section \ref{sec:remarks}. We would like to have a better understanding of
the appearance of the holomorphic anomaly equations; we would like to be
able to calculate the one-loop contribution to the partition function
directly, and we would like to have a cure for the ``missing five-brane''
problem. Besides this, it would of course be interesting to apply our
results to more examples, and in particular to a Calabi-Yau manifold that
has a full $SU(3)$ holonomy group, such as the quintic.

\sk {\bf Acknowledgements.} We would like to thank David Berman for
explaining the correct spelling of the word ``five-brane'' to us. We
thank Jan de Boer, Dan Freed, Greg Moore, Cumrun Vafa, Bernard de Wit,
and Edward Witten for useful discussions. R.D. would like to thank the
Institute for Advanced Study in Princeton, and M.V. would like to
thank the Spinoza Institute and Princeton University for hospitality
during this work. This research was supported by the FOM/NWO Programme
{\it Mathematical Physics}. E.V. is supported by DOE grant
DE-FG02-91ER40571.

\appendix
\newsection{Derivation of the BCOV-equations}
\label{app:bcov}
In section \ref{sec:anomalyeq}, we obtained two equations for the 
conformal blocks $\Psi(x^I; z^I, \zbar^I)$:
\bea
 \label{eq:der1}
 \d_I \Psi & = & \left[ \d_I^x -{i\over 2} C_{IJ}{}^K x^J \d_K^x - i \pi 
 C_{IJK} x^J x^K \right] \Psi, \\
 \label{eq:der2}
 ( \delbar_{\Ibar} - \delbar_{\Ibar} f_1)\Psi & = & \left[{1 
 \over 16\pi i} \Cbar_{\Ibar}{}^{JK} \d_J^x \d_K^x \right]\Psi.
\eea
Throughout this appendix we will use a shorhand for derivatives where 
$\d_I, \delbar_{\Ibar}$ are derivatives with respect to $z^I$ and 
$\zbar^I$ respectively, and derivatives with respect to another variable 
such as $x^I$ have this variable as a superscript.

\sk
We want to relate these equations to the holomorphic anomaly equations for 
the topological string partition function $\Psihat(x^i; z^i, \zbar^i; 
\gl)$, that were found \cite{Bershadsky:1994cx} by Bershadsky, Cecotti, 
Ooguri and Vafa (BCOV). These equations read
\bea
 \d_i \Psihat & = & \left[ \d^x_i - \Gamma_{ij}^k x^j \d_k^x - 
 \frac{1}{2 \gl^2} C_{ijk} x^j x^k + \d_i K \gl \d^\gl \right] \Psihat
 \ret
 \left( \delbar_{\ibar} - \delbar_{\ibar} f_1 \right) \Psihat & = & 
 \left[ \frac{\gl^2}{2} \Cbar_{\ibar\jbar\kbar} e^{2K} G^{j \jbar} G^{k 
 \kbar} \d_j^x \d_k^x - G_{\ibar j} x^j (x^k \d^x_k + \gl \d^\gl) \right] 
 \Psihat.
 \label{eq:bcov}
\eea
The equations look quite similar, but there are three important 
differences, which will be discussed below.

\sk
\vspace{2 mm}
{\em Dependence on $\gl$}

\sk
First of all, the BCOV-equations contain the string coupling constant 
$\gl$. In our approach, we set the prefactor of $1/\gl^2$ in the action to 
1; we can reintroduce it by scaling $x$ and $z$ by a factor of $\gl$. Our 
equations now obtain the same factors of $\gl$ as (\ref{eq:bcov}).

\sk
Reintroducing $\gl$ means that our result will be a homogeneous function of 
degree zero in $z, x,$ and $\lambda$. This gives us an extra differential 
equation:
\be
 \label{eq:hom1}
 \left(\lambda{\d\over\d\lambda}+x^I{\d\over\d x^I}+z^I{\d\over\d z^I} 
 \right)\Psi=0.
\ee

\sk
\vspace{2mm}
{\em From large to small phase space}

\sk
Our equations are written in terms of the coordinates $z^I$, living in 
the ``large phase space'' $H^{3,0}(X) \oplus H^{2,1}(X)$. However, the 
original BCOV-equations are written in terms of the space of all complex 
structures on $X$, which is locally isomorphic to $H^{2,1}(X)$. Therefore, 
we want to restrict our differential equations to some 
surface\footnote{The reader should not be confused by the similar notation 
for the completely different quantities $f_1$ and $f^I$.} $z^I = 
f^I(z^i)$, where the number of $i$ is one less than the number of 
$I$, and the surface cuts every line through the origin in $z^I$-space 
once.

\sk
We want to do something similar to $x$, but at first it is not obvious how 
we should choose the $x^i$ to get to the BCOV-equations. However, 
it is clear from the form of the conformal blocks (\ref{eq:confblock}) 
that the part $x_0^I$ of $x^I$ in the direction of $z^I$ can be absorbed 
in $z^I$. Our equation therefore only depends on the perpendicular part 
$x^I_\bot$ of $x^I$. This quantity depends on $z^I$, so if we define 
$\Psihat(x; z, \zbar; \gl) = \Psi(x_\bot; z, \zbar; \gl)$, its 
$z$-derivatives are given by
\be
 \d_I \Psihat = \left(\d_I + \frac{\d x_\bot^J}{\d z^I} \d_J^x \right) 
 \Psi,
 \label{eq:psihatder}
\ee
and similarly for the $\zbar$-derivatives.

\sk
\vspace{2 mm}
{\em The K\"ahler metric}

\sk
Finally, we have to express our equations in terms of the K\"ahler 
(Zamolodchikov) metric defined by $e^{-K} = \int \Omegabar \wedge \Omega$, 
which in our notation becomes
\be
 K = - \log (z^I \zbar_I),
 \label{eq:kahler}
\ee
where we ignored a constant term.

\sk
An important property of $K$ is that it is constant on surfaces of 
constant $|z|$, so we can use its first derivative to project vectors and 
vector fields in the $z$-direction. For example, we have
\be
 x_\bot^I = x^I + \d_J K z^I x^J,
\ee
which allows us to calculate the derivatives appearing in 
(\ref{eq:psihatder}):
\bea
 \d_I x_\bot^J & = & (\gd_I^J + \d_I K z^J) \d_L K x^L = 0 \\
 \label{eq:xderz}
 \delbar_{\Ibar} x_\bot^J & = & G_{\Ibar K} x^K z^J,
\eea
where in the first line we used that $\d_L K x^L = \d_L K x_0^L$ which 
vanishes since we set $x_0 = 0$, and in the second line $G_{\Ibar J}$ is 
the second derivative of the K\"ahler form $K$.

\sk
\vspace{2 mm}
{\em Some more relations}
\nopagebreak

There are a few more relations we have to derive before we can rewrite 
our equations. First, note that $z^I C_{IJK} = 0$, which we can insert in 
(\ref{eq:der1}) and (\ref{eq:der2}) to obtain
\bea
 \label{eq:hom2}
 z^I {\d\over \d z^I}\Psi & = & z^I {\d\over\d x^I} \Psi \\
 \label{eq:hom3}
 \zbar^{I} \delbar_{\Ibar} \Psi & = & 0.
\eea
The relations (\ref{eq:hom1}), (\ref{eq:hom2}) and (\ref{eq:hom3}) will be the 
key ingredients in rewriting our equations in terms of the new variables.

\sk
Finally, we would like to express the derivatives $\d_i$ in terms of 
$\d_I$. Note that
\bea
 \frac{\d \Psi}{\d z^i}
 & = & \left( \frac{\d f^I_\bot}{\d z^i} + \frac{\d f^I_0}{\d z^i} 
       \right) \frac{\d \Psi}{\d z^I} \ret
 & = & \left( \frac{\d f^I_\bot}{\d z^i} - \frac{\d K}{\d z^i} z^I \right) 
       \frac{\d \Psi}{\d z^I},
 \label{eq:smallder}
\eea
where in the last line we used the fact that $f_0^I$ is uniquely 
parameterized by $K$. Of course, everything we did here also holds for the 
$\zbar^I$-derivatives.

\sk
\vspace{2 mm}
{\em The BCOV-equations}

\sk
So now we collect everything we derived so far to reach our final answer. 
Using (\ref{eq:smallder}), (\ref{eq:hom1}) and (\ref{eq:der1}) we obtain
\be
 \d_i \Psihat = \d_i f^I_\bot \left( \d^x_I - \frac{i}{2} {C_{IJ}}^K 
 x^J_\bot \d^x_K - \frac{1}{2 \gl^2} C_{IJK} x^J_\bot x^K_\bot \right) 
 \Psi + \d_i K (\gl \d^\gl + x^I_\bot \d_I^x) \Psi.
 \label{eq:delpsi}
\ee
Now it is clear what is the right way to define the $x^i$: they should be 
such that
\be
 x^I_\bot = \frac{\d f^I_\bot}{\d z^i} x^i.
\ee
Inserting this in (\ref{eq:delpsi}), we can remove all upper case indices 
and obtain
\bea
 \d_i \Psihat 
 & = & \left( \d^x_i - \frac{i}{2} {C_{ij}}^k x^j \d^x_k - 
       \frac{1}{2 \gl^2} C_{ijk} x^j x^k \right) \Psihat + \d_i K (\gl \d^\gl + 
       x^i \d_i^x) \Psihat.
\eea
Now note that the K\"ahler metric and the metric $\Im \tau$ are related by
\be
 \Im \tau_{IJ} = - 2iG_{\Ibar J} e^{-K},
\ee
when acting on the perpendicular coordinates. From this equation, it is 
straightforward to write down the Cristoffel symbols for the K\"ahler 
metric:
\be
 \gG_{ijk} = \half \d_i G_{jk} = ( \frac{1}{8} C_{ijk} + \frac{i}{4} \d_i K (\Im 
 \tau)_{jk} ) e^K.
\ee
Inserting this in our result gives:
\be
 \d_i \Psihat = \left[ \d^x_i - \Gamma_{ij}^k x^j \d_k^x - 
 \frac{1}{2 \gl^2} C_{ijk} x^j x^k + \d_i K \gl \d^\gl \right] \Psihat,
\ee
which is the first BCOV-equation.

\sk
The calculations to get to the second BCOV-equation are very similar: 
using the definitions and the homogeneity relations, we find
\be
 \delbar_{\ibar} \Psihat = \delbar_{\ibar} f^I_\bot \left( \delbar_{\Ibar} 
 f_1 + \frac{\gl^2}{8} {\Cbar_{\Ibar}}^{JK} \d_J^x \d_K^x - G_{\Ibar K} 
 x_\bot^K (x_\bot^J \d^x_J + \gl \d^\gl) \right) \Psi.
\ee
Now we can again insert the K\"ahler metric and replace upper case indices 
by lower case ones, and we obtain the second BCOV-equation:
\be
 \left( \delbar_{\ibar} - \delbar_{\ibar} f_1 \right) \Psihat = \left[
 \frac{\gl^2}{2} \Cbar_{\ibar \jbar \kbar} e^{2K} G^{j \jbar} G^{k \kbar} 
 \d_j^x \d_k^x - G_{\ibar j} x^j (x^k \d^x_k + \gl \d^\gl) \right] 
 \Psihat,
\ee

\end{document}